\documentclass[preprint,12pt]{elsarticle}

\usepackage{amsmath,amssymb}
\usepackage{graphicx}
\begin{document}

\begin{frontmatter}

\title{Demodulation of a positron beam in a bent crystal channel}

\author[fias]{A.~Kostyuk}
\ead{kostyuk@fias.uni-frankfurt.de}
\author[fias,smtu]{A.V.~Korol}
\ead{a.korol@fias.uni-frankfurt.de}
\author[fias]{A.V.~Solov'yov}
\ead{solovyov@fias.uni-frankfurt.de}
\author[fias]{Walter Greiner}

\address[fias]{Frankfurt Institute for Advanced Studies,
 Johann Wolfgang Goethe-Universit\"at, 
Ruth-Moufang-Str.~1, 60438 Frankfurt am Main, Germany}
\address[smtu]{Department of Physics,
St Petersburg State Maritime Technical University,
St Petersburg, Russia}

\begin{abstract}
The evolution of a modulated positron beam in a planar crystal channel
is investigated within the diffusion approach. A detailed description 
of the formalism is given.
A new parameter, the demodulation length, is introduced, representing the quantitative 
measure of the depth at which the channelling beam preserves its modulation in the crystal.
It is demonstrated that there exist crystal channels with the demodulation length
sufficiently large for using the crystalline undulator as a coherent source of hard X rays.
This finding is a crucial milestone in developing a new type of lasers radiating
in the hard X ray and gamma ray range.
\end{abstract}

\begin{keyword}

\PACS 61.85.+p \sep 05.20.Dd \sep 41.60.-m


\end{keyword}

\end{frontmatter}

%\maketitle
%---------------------------------------------------------------------------

\section{Introduction}
In this article we study the evolution of a modulated 
positron beam in straight and bent planar crystal channels.
Some key ideas of this research were briefly communicated in \cite{demod_jpg}
and \cite{demod_Chan2008}. In this paper we present a systematic and detailed description 
of the formalism and the obtained results.
The outcome of the research is of crucial importance for
the theory of the crystal undulator based laser (CUL) \cite{first,KSG1999,klystron}
--- a new electromagnetic radiation source in hard x- and gamma-ray range.

Channelling takes place if charged particles enter a single crystal at small
angle with respect to crystallographic planes or axes \cite{Lindhard}. The 
particles get confined
by the interplanar or axial potential and follow the shape of the corresponding 
planes and axes. This suggested the idea \cite{Tsyganov1976} of using bent crystals
to steer the particles beams. Since its first experimental verification \cite{Elishev1979}
the idea to deflect or extract  high-energy charged particle beams by means of tiny bent crystals
replacing huge dipole magnets has been attracting a lot of interest worldwide. Bent crystal
have been routinely used for beam extraction in the Institute for High Energy Physics, Russia
\cite{Afonin2005}. A series of experiments on the bent crystal deflection of proton and heavy ion beams was
performed at different
accelerators \cite{Arduini1997,Scandale2008,Carrigan1999,Fliller2006,Strokov2007} throughout the world.
The bent crystal method has been proposed to extract particles from the beam halo at
CERN Large Hadron Collider \cite{Uggerhoj2005}
The possibility of deflecting positron \cite{Bellucci2006} and electron \cite{Strokov2007,Strokov2006}
beams has been studied as well.

\begin{figure}[htb]
%\onefigure[width=8.5cm]{undulator.eps}
\begin{center}
\includegraphics*[width=12cm]{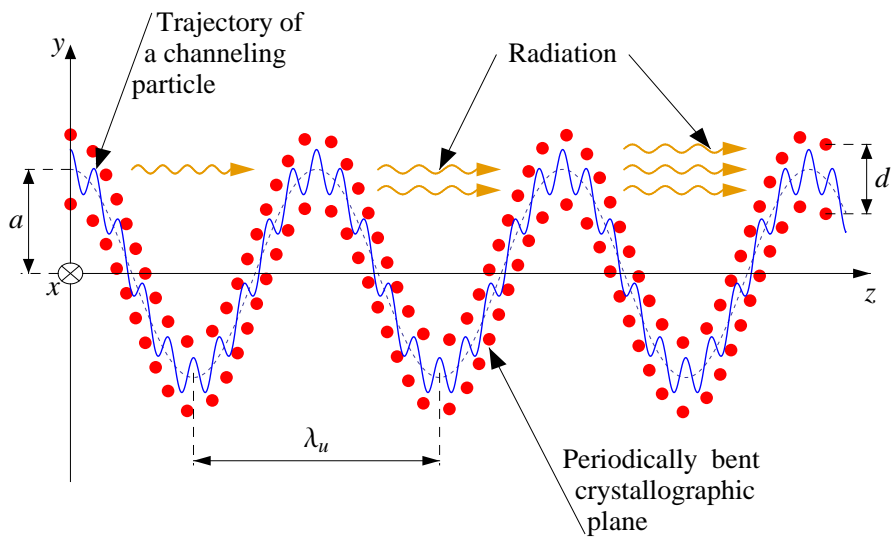}
\end{center}
%%% for two column
%\includegraphics*[width=8.5cm]{undulator.eps}
\caption{Schematic representation of the crystalline undulator.
%Circles denote the atoms belonging to 
%neighboring crystallographic planes (separated by the distance $d$)
%which are periodically bent.
%The wavy lines represent the trajectories of channelling particles.
}
\label{undulator.fig}
\end{figure}

A single crystal with {\it periodically} bent crystallographic planes can force 
channelling particles to move along  nearly sinusoidal trajectories
and radiate in the hard x- and gamma-ray frequency range (see figure \ref{undulator.fig}).
The feasibility of such a device, known as the 'crystalline undulator`,
was demonstrated theoretically a decade ago \cite{first} (further
developments as well as historical references are reviewed in \cite{KSG2004_review}).
More recently, an electron based crystalline undulator has been proposed 
\cite{Tabrizi}.

It was initially suggested to obtain sinusoidal bending
by the propagation of an acoustic wave along the crystal 
\cite{first,KSG1999}. The advantage of this approach is its flexibility:
the period of deformation can be chosen by tuning the frequency of the ultrasound.
However, this approach is rather challenging technologically and yet to be tested 
experimentally.
Several other technologies for the  manufacturing of periodically bent crystals
have been developed and tested. These include
making regularly spaced grooves on the crystal surface
either by a diamond blade \cite{BellucciEtal2003,GuidiEtAl_2005}
or by means of laser-ablation \cite{Balling2009},
deposition
of periodic
Si$_3$N$_4$ layers onto the surface of a Si crystal \cite{GuidiEtAl_2005},
growing of Si$_{1-x}$Ge$_x$ crystals\cite{Breese97}
with a periodically varying Ge content $x$ \cite{MikkelsenUggerhoj2000,Darmstadt01}.

Experimental studies of the crystalline undulator are currently in progress. The first results 
are reported in \cite{Baranov2006} and \cite{Backe2008}.

The advantage of the crystalline undulator is in extremely strong
electrostatic fields inside a crystal which are able
to steer the particles much more effectively than even the most advanced
superconductive magnets. 
This fact allows to make the period $\lambda_\mathrm{u}$ of the crystalline undulator
in the range of hundreds or tens micron which is two to three orders of 
magnitude smaller than that of conventional undulator. Therefore
the wavelength of the produced radiation 
$\lambda \sim \lambda_\mathrm{u}/(2 \gamma^2)$ ($\gamma \sim 10^3$--$10^4$ being the 
Lorentz factor of the particle) can reach the (sub)picometer range, 
where conventional sources with 
comparable intensity are unavailable \cite{Topics}.

\begin{figure}[tb]
\begin{center}
\includegraphics*[width=12cm]{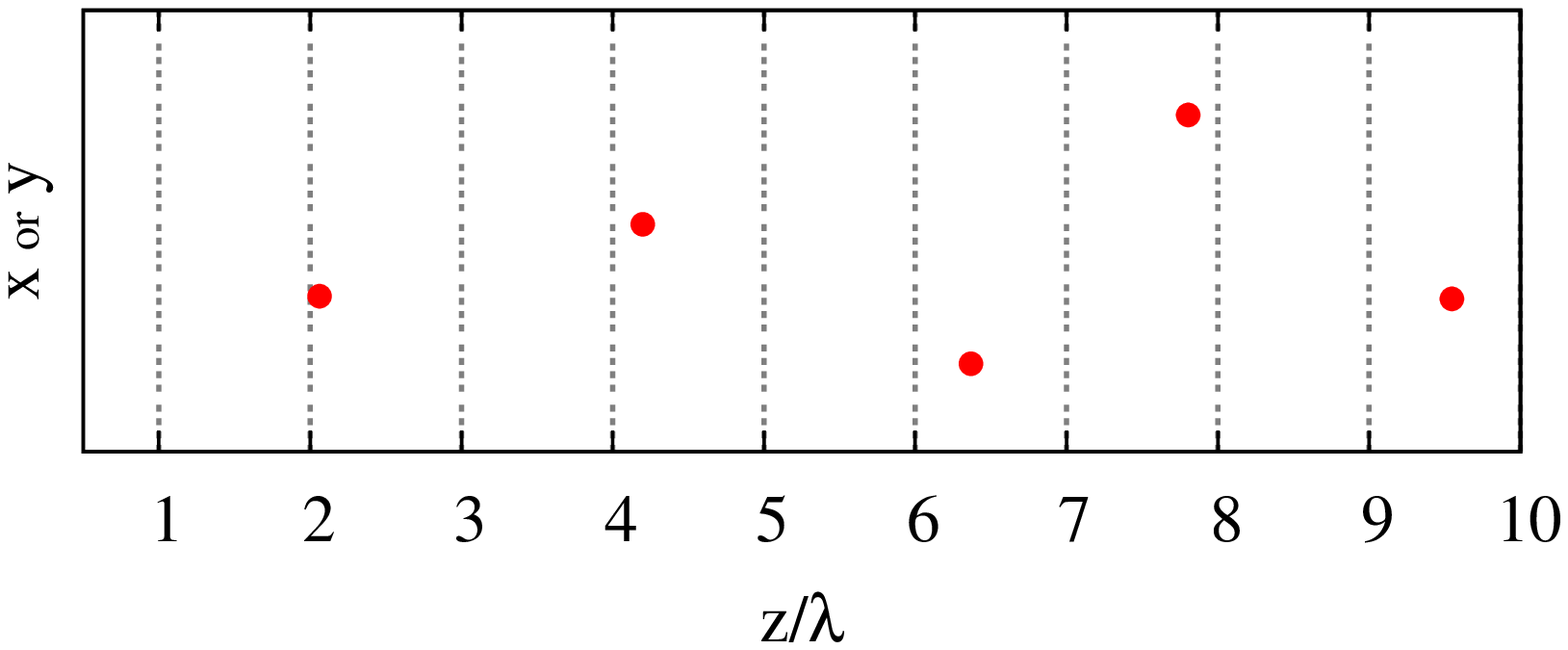}
\includegraphics*[width=12cm]{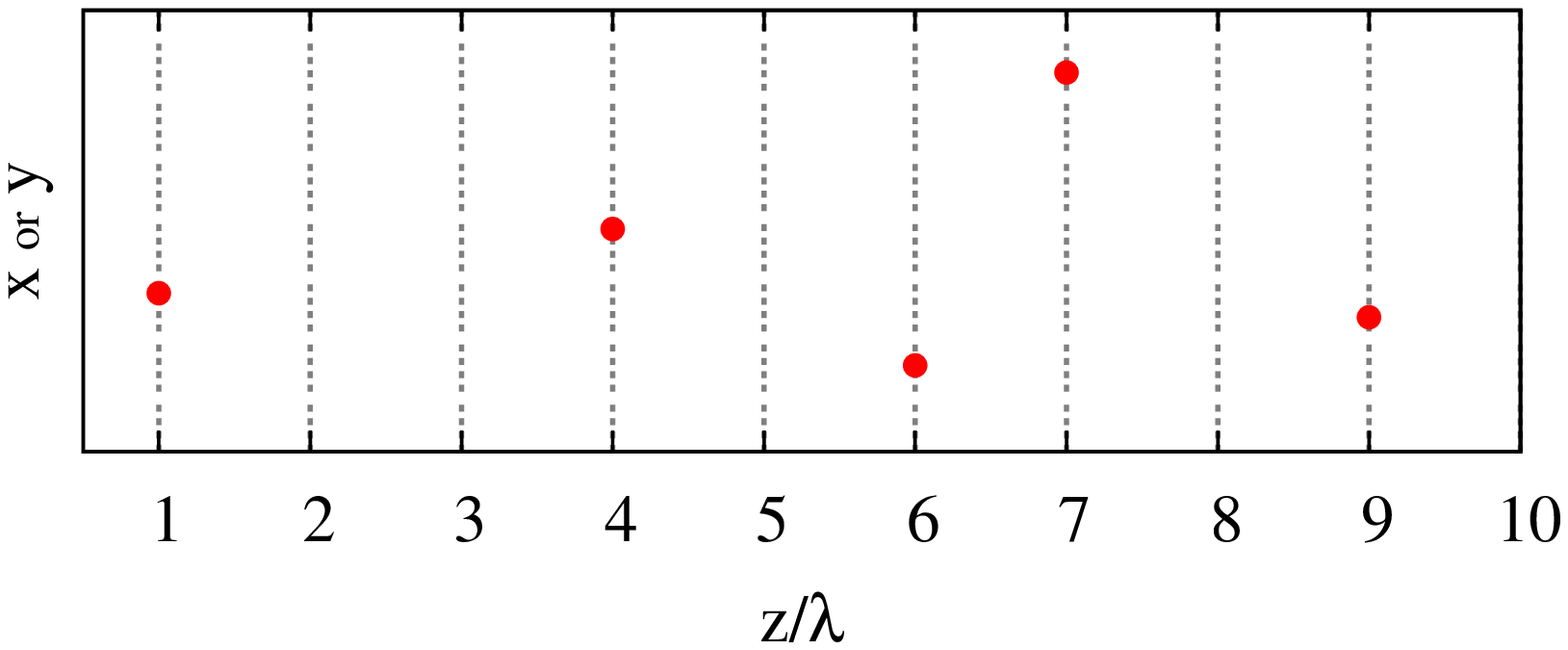}
\end{center}
\caption{In an unmodulated beam (the upper panel) the particles 
are randomly distributed. In a completely modulated beam 
(the lower panel) the distance between any two particles along 
the beam direction is an integer multiple of the modulation 
period $\lambda$.
}
\label{modulation.fig}
\end{figure}

Even more powerful and coherent radiation will be emitted if
the probability density of the particles in the beam is modulated
in the longitudinal direction with the period $\lambda$, equal
to the wavelength of the emitted radiation
(see figure \ref{modulation.fig}).
In this case, the electromagnetic waves emitted  in the forward direction by
different particles have approximately the same phase \cite{Ginzburg}. Therefore,
the intensity of the radiation becomes proportional
to the beam density squared (in contrast to the linear proportionality for an unmodulated
beam). This increases the photon flux {\it by orders of magnitude}
relative to the radiation of unmodulated beam of the same density.
The radiation of a modulated beam in an undulator is 
a keystone of the physics of free-electron lasers (FEL) \cite{Madey,SchmueserBook}.
It can be considered as  a
classical counterpart of the stimulated  emission in quantum physics.
Therefore, if similar phenomenon takes place in a crystalline undulator, it can be
referred to as the {\it lasing regime of the crystalline undulator}.

The feasibility of CUL radiating in the hard
x-ray and gamma-ray range was considered for the fist time in \cite{first,KSG1999}. 
Recently,
a two-crystal scheme, the gamma klystron, has been proposed \cite{klystron}.

\begin{figure}[htb]
%%% for two column
%\includegraphics[width=8.5cm]{DebunchY.eps}
%\includegraphics[width=8.5cm]{DebunchX.eps}
\begin{center}
\includegraphics[width=12cm]{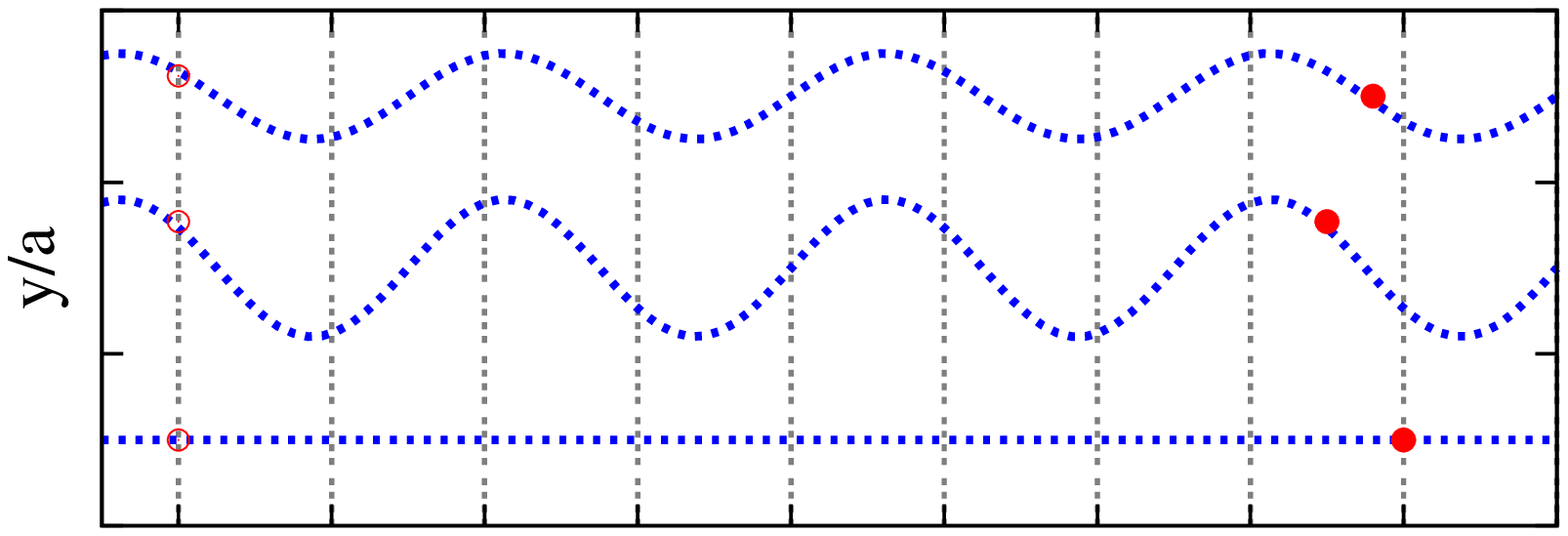}
\includegraphics[width=12cm]{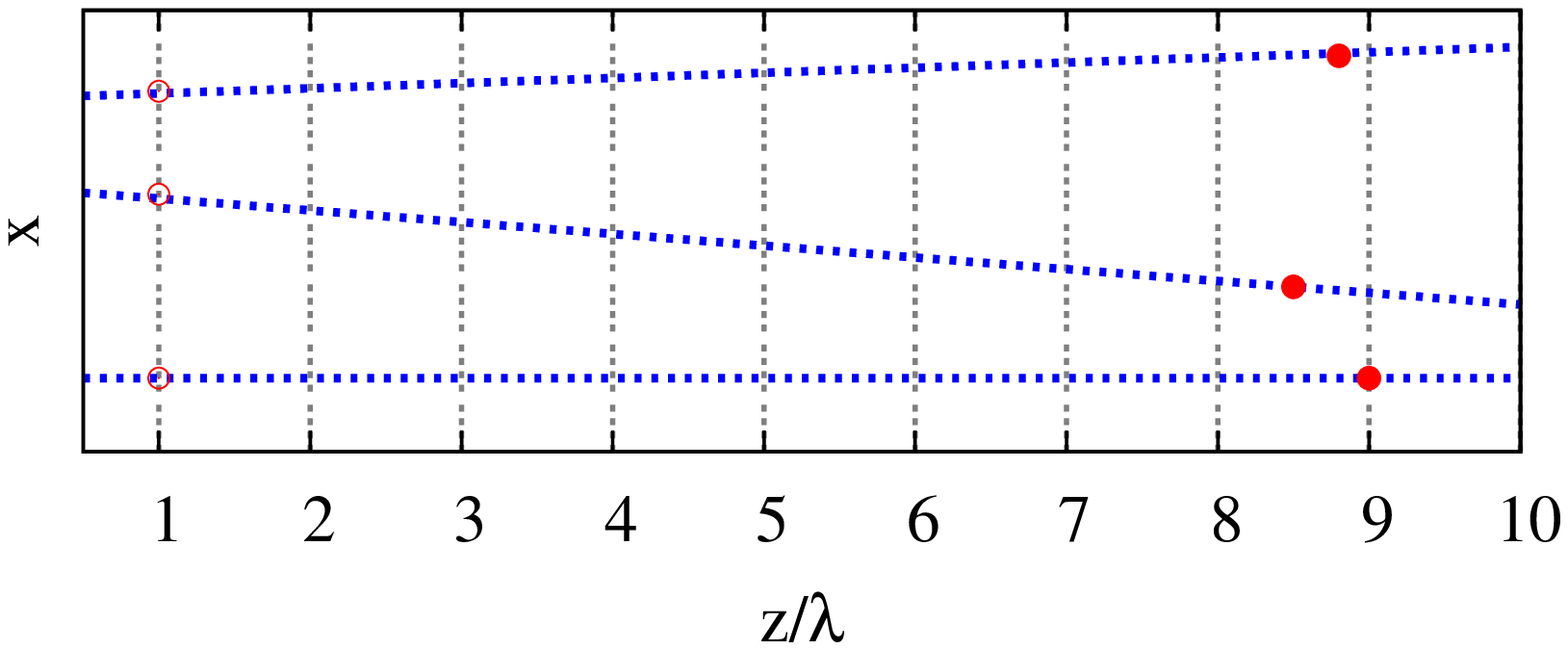}
\end{center}
\caption{Due to different amplitudes of channelling oscillation
(upper panel) and different momentum directions 
in the $(xz)$ plane (lower panel), the initially modulated
beam gets demodulated. The open and filled circles denote the
same particles at the crystal entrance and after
travelling some distance in the crystal channel, respectively.
}
\label{demodulation.fig}
\end{figure}

A simplified model used in the cited papers assumed that all particle trajectories
follow exactly the shape of the bent channel. In reality, however, the particle
moving along the channel also oscillates in the transverse direction
with respect 
to the channel axis (see the shape of the trajectory 
in figure \ref{undulator.fig}). Different particles have
different amplitudes of the oscillations inside the channel
(figure \ref{demodulation.fig}, upper panel).
Similarly, the directions of particle momenta
in $(xz)$ plane are slightly different
(figure \ref{demodulation.fig}, lower panel).
Even if the speed of the particles along their
trajectories is the same, the particles oscillating with 
different amplitudes or the particles with different trajectory
slopes with respect to $z$ axis have slightly
different components of their velocities along the channel.
As a result, the beam gets demodulated.
An additional contribution to the beam demodulation comes 
from incoherent collisions
of the channelling particles with the crystal constituents.

In the case of an unmodulated beam, 
the length of the crystalline undulator 
and, consequently, the maximum accessible intensity of the radiation
are limited by the dechannelling process.
The channelling particle gradually gains the energy of transverse 
oscillation due to collisions with crystal constituents.
At some point this energy exceeds the maximum value of the interplanar
potential and the particle leaves the channel. The average penetration length 
at which this happens is known as the {\it dechannelling length}.
The dechannelled particle no longer follows the sinusoidal 
shape of the channel and, therefore, does not contribute to the 
undulator radiation. Hence, 
the reasonable length of the crystalline undulator 
is limited to a few dechannelling lengths. A longer crystal would attenuate rather
then produce the radiation. Since the intensity of the undulator radiation is
proportional to the undulator length squared, the dechannelling length 
and the attenuation length are the 
main restricting factors that have to be taken into account when the radiation output 
is calculated.

In contrast, not only the shape of the trajectory but also the particles positions 
with respect to each other along $z$ axis are important for the lasing regime.
If these positions become random because of the beam demodulation, the intensity
of the radiation drops even if the particles are still in the channelling mode.
Hence, it is the beam demodulation rather than dechannelling that restricts the 
intensity of the radiation of CUL.
Understanding this process and estimating the characteristic length at which this phenomenon
takes place is, therefore, a cornerstone of the theory of this
new radiation source.

\section{Diffusion Equation} 

\subsection{The model of the crystal channel}

\begin{figure}[ht]
%\includegraphics*[width=12cm]{figure1.eps}
%%% for two column
\begin{center}
\includegraphics*[width=10cm]{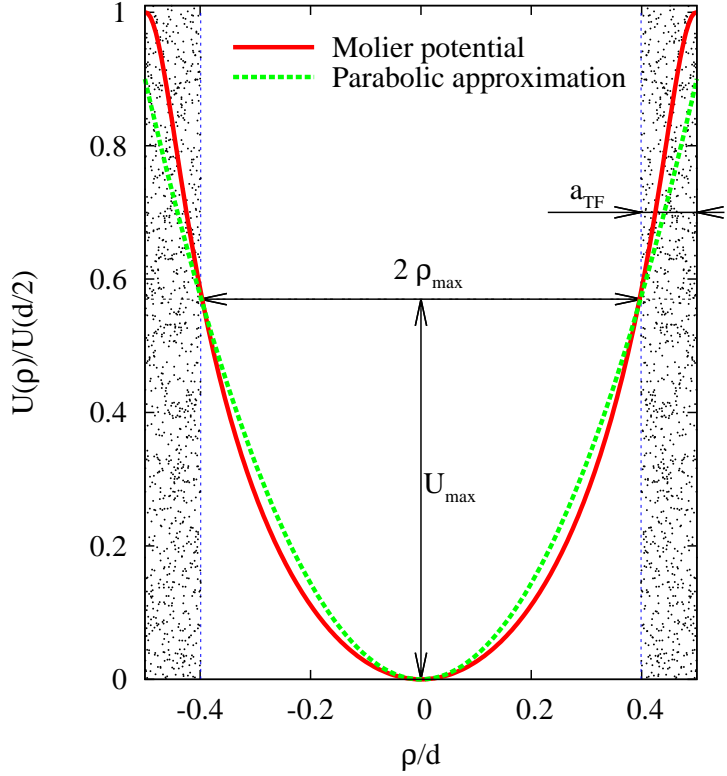}
\caption{The model of the plane crystal channel.
The interplanar potential is approximated by a parabola.
It is assumed that the particle dechannels if it enters the 
vicinity of the crystallographic plane
within the Thomas-Fermi radius, $a_\mathrm{TF}$.
}
\end{center}
\label{potential.fig}
\end{figure}

We adopt the following model of the planar crystal channel (see Fig. \ref{potential.fig}):\\
- the interplanar potential is approximated with a parabola 
\begin{equation}
U(\rho) = U_{\rm max} \left( \frac{\rho}{\rho_{\rm max}} \right)^2
\label{potential}
\end{equation}
($\rho$ is the distance from the potential minimum) 
   so that the channeling oscillations are assumed to be harmonic;\\
- the electron density within the distance of one Thomas-Fermi radius of the crystal atoms 
  from the crystallographic plane is assumed to be so high that the particle gets
  quickly scattered out of the channel. Therefore, the particle is considered dechanneled just after 
  it enters this region. So that the effective channel width is $2 \rho_{\max} = d - 2 a_\mathrm{TF}$,
  where $d$ and $a_\mathrm{TF}$ are respectively the interplanar distance and  is the Thomas-Fermi 
  radius.

As is seen from the figure, the parabolic approximation is quite reasonable. The real potential differs from the 
parabola mostly in the region of high electron density, where the particle assumes to be dechanneled.

\subsection{The particle distribution} 

Let us consider the distribution 
$f(t,s;\xi,E_{y})$
of the beam particles 
with respect to the angle between the particle trajectory
and axis $z$ in the $(xz)$ plane 
$\xi = \arcsin p_{x}/p \approx p_{x}/p$ and the energy of the channeling
oscillation $E_{y} = p_{y}^2/2 E + U(y)$ \footnote{We chose the system of units in such 
a way that the speed of light is equal to unity.
Therefore, mass, energy and momentum have the same dimensionality. This is also true
for length and time.}. Here $p$, $p_{x}$ and $p_{y}$ are, respectively, the particle momentum and its $x$ and $y$ components, and 
$E$ is the particle energy (we will consider only ultrarelativistic 
particles, therefore $E \approx p$).

\subsection{Kinetic equation} 

In absence of random scattering, the distribution function $f(t,z;\xi,E_{y})$
would satisfy the differential equation 
$ \frac{\partial f}{\partial t} +  
\frac{\partial f}{\partial z} v_{z}
= 0$,
where $v_{z} \equiv \frac{\partial z}{\partial t}$.
In reality, however, the right-hand-side of the equation is not zero. It contains the
collision integral. After averaging over the period  of the channeling oscillation, the kinetic 
equation takes the form
\begin{eqnarray}
\frac{\partial f}{\partial t} +  
\frac{\partial f}{\partial z} \langle v_{z} \rangle
& = & 
\left \langle
\int \! \! \! \! \! \int \! d \xi d \! E_{y}'  
\left[ 
f(t,z;\xi',E_{y}') w(\xi',E_{y}'; \xi, E_{y}) 
\right. \right.  \nonumber \\
& & 
\left. \left.
- f(t,z;\xi,E_{y}) w(\xi,E_{y}; \xi', E_{y}') 
\right ] 
\right \rangle
\end{eqnarray}
where $w(\xi,E_{y}; \xi', E_{y}') dz$ is the probability that the particle changes its angle $\xi$ and 
transverse energy from $\xi$ and $E_{y}$ to, respectively, $\xi'$ and $E_{y}'$
while travelling the distance $dz$.
The angular brackets stand for averaging over the period of the channeling oscillations.

Due to the detailed equilibrium 
\begin{equation}
w(\xi,E_{y}; \xi', E_{y}') = w(\xi',E_{y}'; \xi, E_{y})
\end{equation}

\subsection{Diffusion approximation} 

We assume that soft scattering dominates, i.e. the function $w(\xi',E_{y}'; \xi, E_{y})$ is not negligible 
only if $|\xi' - \xi|$ and $|E_{y}' - E_{y}|$ are small so that 
$f(t,z;\xi',E_{y}') \equiv f(t,z;\xi + \vartheta_{x},E_{y}+q_{y})$ 
can be expanded into the Taylor series with respect to $\vartheta_{x}$ and $q_{y}$.
Then, up to the second order in $\vartheta_{x}$ and $q_{y}$, one obtains
\begin{equation}
\frac{\partial f}{\partial t} +  
\frac{\partial f}{\partial z} \langle v_{z} \rangle
 = D_{\xi} \frac{\partial f}{\partial \xi}
+ D_{y} \frac{\partial f}{\partial E_{y}}
+ D_{\xi \xi}  \frac{\partial^2 f}{\partial \xi^2}
+ D_{\xi y} \frac{\partial^2 f}{\partial \xi \partial E_{y}}
+ D_{y y} \frac{\partial^2 f}{\partial  E_{y}^2}
\label{difeqD}
\end{equation}
where
\begin{eqnarray}
D_{\xi}      \! &\! = \!& \! \left \langle \int d \vartheta_{x} \, \vartheta_{x}   \int d q_{y} 
w(\xi,E_{y};\xi + \vartheta_{x},E_{y} + q_{y}) \right \rangle  \label{Dxi} \\
D_{y}        \! &\! = \!& \! \left \langle \int d \vartheta_{x}  \int d q_{y} \, q_{y}  
w(E_{\xi},E_{y};\xi + \vartheta_{x},E_{y} + q_{y}) \right \rangle  \label{Dy} \\
D_{\xi \xi}  \! &\! = \!& \! \frac{1}{2} \left \langle \int d \vartheta_{x} \, \vartheta_{x}^2   \int d q_{y} 
w(\xi,E_{y};\xi + \vartheta_{x},E_{y} + q_{y}) \right \rangle  \label{Dxx} \\
D_{\xi y}    \! &\! = \!& \! \left \langle \int d \vartheta_{x} \, \vartheta_{x}   \int d q_{y}  \, q_{y}
w(\xi,E_{y};\xi + \vartheta_{x},E_{y} + q_{y}) \right \rangle  \label{Dxy} \\
D_{yy}       \! &\! = \!& \! \frac{1}{2} \left \langle \int d \vartheta_{x}  \int d q_{y} \, q_{y}^2  
w(\xi,E_{y};\xi + \vartheta_{x},E_{y} + q_{y}) \right \rangle  \label{Dyy}  
\end{eqnarray}

\section{Diffusion Coefficient} 

\subsection{Scattering}

Let us consider a channeling positron colliding with a target electron. If $\theta$ is the scattering angle 
in the lab frame and $\varphi$ is the angle between the scattering plane and the $(xz)$-plane
then the transverse components of the particle momentum are changed by 
\begin{eqnarray}
\delta p_x &=& p \sin \theta \cos \varphi, \\
\delta p_y &=& p \sin \theta \sin \varphi, 
\end{eqnarray}
As far as $\theta \ll 1$, we can use the approximation $\sin \theta \approx \theta$.
Then 
\begin{equation}
\vartheta_{x} = \frac{p_x + \delta p_x}{p} - \frac{p_x}{p} = \frac{\delta p_x}{p} 
=  \theta \cos \varphi .
\end{equation}
and
\begin{eqnarray}
q_{y} &=& \left( \frac{(p_y + \delta p_y)^2}{2 E} + U(y) \right) - \left( \frac{p_y^2}{2 E} + U(y) \right) \\
&=&  p_y \theta \sin \varphi + 
\frac{p}{2} \theta^2 \sin^2 \varphi.
\nonumber
\end{eqnarray}

\subsection{The transition probability}

The probability for the particle to be scattered by an electron from the 
state $(\xi,E_{y})$ to the state $(\xi+\vartheta_{x},E_{y}+q_{y})$  while travelling the 
distance $dz$ can be related to the differential cross section of positron-electron scattering: 
\begin{eqnarray}
w(\xi,E_{y};\xi + \vartheta_{x},E_{y} + q_{y}) d z = 
n_e d z \int d \theta \int_{0}^{2 \pi} d \varphi \frac{d^2 \sigma}{d \theta  d \varphi}   
\hspace*{1em}
& & 
\label{wviasigma}\\
\delta \left (  \theta \cos \varphi - \vartheta_{x} \right ) \ 
\delta \left ( p_y \theta \sin \varphi + 
\frac{p}{2} \theta^2 \sin^2 \varphi - q_{y} \right ) & & 
\nonumber
\end{eqnarray} 
Because both target and projectile are not polarized, the cross section does not 
depend on $\varphi$:
\begin{equation}
\frac{d^2 \sigma}{d \theta  d \varphi} = \frac{1}{2 \pi} \frac{d \sigma}{d \theta} .
\end{equation}

Substituting (\ref{wviasigma}) into (\ref{Dxi})--(\ref{Dyy})
and integrating over $\vartheta_{x}$ and $q_{y}$ one obtains
\begin{eqnarray}
D_{\xi} &=& \frac{1}{2 \pi} \left \langle n_e \int d \theta
\int_{0}^{2 \pi} d \varphi   \frac{d \sigma}{d \theta} \theta \cos \varphi
 \right \rangle \\
D_{y} &=& \frac{1}{2 \pi} \left \langle n_e \int d \theta 
\int_{0}^{2 \pi} d \varphi  \frac{d \sigma}{d \theta} 
%\right.
%\\
%& & 
%\left.
%\hspace{4em}
\left( 
p_y \theta \sin \varphi + 
\frac{p}{2} \theta^2 \sin^2 \varphi
\right)  
 \right \rangle 
%\nonumber 
\\
D_{\xi \xi} &=& \frac{1}{4 \pi} \left \langle n_e \int d \theta 
\int_{0}^{2 \pi} d \varphi  \frac{d \sigma}{d \theta}
 \theta^2 \cos^2 \varphi 
 \right \rangle \\
D_{\xi y} &=& \frac{1}{2 \pi} \left \langle n_e \int d \theta \frac{d \sigma}{d \theta}
\int_{0}^{2 \pi} d \varphi  
 \theta \cos \varphi 
%\right. \\
%& & 
%\hspace{4em}
%\left .
\left( 
p_y \theta \sin \varphi + 
\frac{p}{2} \theta^2 \sin^2 \varphi
\right) 
 \right \rangle 
%\nonumber 
\\ 
D_{yy} &=& \frac{1}{4 \pi} \left \langle n_e \int d \theta \frac{d \sigma}{d \theta}
\int_{0}^{2 \pi} d \varphi  \theta^2 \sin^2 \varphi
%\right. \\
%& & 
%\hspace{4em}
%\left .
\left( p_y  + 
\frac{p}{2} \theta \sin \varphi
\right)^2  
 \right \rangle 
%\nonumber
\end{eqnarray}

Then integration over $\varphi$ and neglecting higher order terms
with respect to $\theta$ yields
\begin{eqnarray}
D_{\xi}  &=& 0 \\
D_{y} &=& \frac{p}{4} \langle n_e \rangle \int d \theta \frac{d \sigma}{d \theta} 
\theta^2 \\
D_{\xi \xi} &=& \frac{1}{4} \langle n_e  \rangle
\int d \theta \frac{d \sigma}{d \theta} \theta^2 \\
D_{\xi y} &=& 0 \\
D_{yy} &=& \frac{1}{4} \langle n_e p_{y}^2 \rangle
\int d \theta \frac{d \sigma}{d \theta} \theta^2
\end{eqnarray}

Here $\langle n_e  \rangle$ is the electron density along the particle trajectory averaged
over the period of the channeling oscillations. Generally speaking, 
$\langle n_e  \rangle$ depends on the transverse energy $E_{y}$
We assume, however, that the electron density does not change 
 essentially within the channel. Therefore, $\langle n_e  \rangle$ can be treated 
as a constant. For the same reason,  we can make  the approximation 
$
\langle n_e p_{y}^2 \rangle \approx  \langle n_e \rangle \langle  p_{y}^2 \rangle
$
Then 
$
\langle p_{y}^2 \rangle = 
2 E \left \langle \frac{p_{y}^2}{2 E} \right \rangle = 
E E_{y}.
$
due to the virial theorem for the harmonic potential: $p_{y}^2/(2 E) =  E_{y}/2$

Finally, one obtains for nonzero coefficients
\begin{eqnarray}
D_{y} &\equiv& D_{0} \\
D_{\xi \xi} &=& \frac{1}{E} D_{0} \\
D_{yy} &=& E_{y} D_{0}
\end{eqnarray}

The diffusion equation takes the form
\begin{equation}
\frac{\partial f}{\partial t} +  
\frac{\partial f}{\partial z} \langle v_{z} \rangle
 = D_0 \left [ \frac{\partial }{\partial E_{y}}
\left(E_{y}
\frac{\partial f }{\partial E_{y}}
\right)
+
\frac{1}{E} 
\frac{\partial^2 f }{\partial \xi^2}
\right ] \ .
\label{diffeqD_0}
\end{equation}
Equation (\ref{diffeqD_0}) is akin to the equation describing 
dechanneling process (see e.g. \cite{BiryukovChesnokovKotovBook}).
The novel feature of it is the presence of time variable, which allows
to describe time dependent (modulated) beams. Additionally, it takes 
into account scattering in the $(x,z)$ plane.

\section{Solving the diffusion equation}
\label{solving}

\subsection{The average longitudinal velocity}

The particle velocity along $z$ axes averaged over the period of 
channeling oscillations can be represented as 
\begin{equation}
\langle v_{z} \rangle = 
\frac{
 \sqrt{1 - \frac{1}{\gamma^2}} \cos \xi
}
{
\frac{k_{\rm c}}{2 \pi}
\int_0^{2 \pi/k_{\rm c}}  
\sqrt{1 + \left [ b k_{\rm c}  \sin(k_{\rm c} z) \right ]^2 } d z
}.
\end{equation} 
Here $\sqrt{1 - 1/\gamma^2}$ is the particle speed along the trajectory,
$\cos  \xi \approx (1 - \xi^2/2)$  appears because of the slope $\xi \ll 1$ of the trajectory to $z$ axis in 
$(xz)$ plane, and the denominator is due to the sinusoidal channeling
oscillations in $(xy)$ plane with the amplitude $b$ and the period $\lambda_{\rm c} = 2 \pi/k_{\rm c}$.
Taking into account that the amplitude of the channeling oscillations is much smaller
than their period, $b k_{\rm c} \ll 1$, the denominator can be approximated by 
$1 + (b k_{\rm c})^2/4$.
For the harmonic potential (\ref{potential}) (see Fig. \ref{potential.fig})
the amplitude $b$ is related to the transverse energy $E_{y}$  by
\begin{equation}
b = \rho_{\rm max} \sqrt{\frac{E_{y}}{U_{\rm max}}}.
\end{equation} 
Using the formula for the frequency of the harmonic oscillator one finds
\begin{equation}
k_{\rm c} =   \sqrt{\frac{1}{E}\frac{d^2 U}{d \rho^2}} = 
\frac{1}{\rho_{\rm max}} \sqrt{\frac{2 U_{\rm max}}{E}}
\end{equation} 
So that 
$
b k_{\rm c} = \sqrt{\frac{2 E_{y}}{E}}.
$
Finally, neglecting higher order terms
\begin{equation}
\langle v_{z} \rangle
 \approx \left( 
1 - \frac{1}{2 \gamma^2} - \frac{\xi^2}{2} - \frac{E_{y}}{2 E}
\right)
\label{v_z}
\end{equation}

\subsection{Excluding the time variable}

If the beam is periodically modulated (bunched) the distribution $f(t,z;\xi,E_{y})$
can be represented as a Fourier series:
\begin{equation}
f(t,z;\xi,E_{y}) = \sum_{j=-\infty}^{\infty} g_{j} (z;\xi,E_{y}) \exp (i j \omega t).
\end{equation}
with $g_{j}^{*} (z;\xi,E_{y}) = g_{-j} (z;\xi,E_{y})$ to ensure the real value of the
particle distribution. Since Eq. (\ref{diffeqD_0}) is linear, 
it is sufficient to consider only one harmonic.
Substituting 
$f(t,z;\xi,E_{y}) = g (z;\xi,E_{y}) \exp (i \omega t)$ into (\ref{diffeqD_0}) one obtains
\begin{equation}
i \omega g(z;\xi,E_{y}) +
 \frac{\partial g}{\partial z} 
\langle v_{z} \rangle
 = D_0 \left [ \frac{\partial }{\partial E_{y}}
\left(E_{y}
\frac{\partial g }{\partial E_{y}}
\right)
+
\frac{1}{E} 
\frac{\partial^2 g }{\partial \xi^2}
\right ] .
\label{diffeqg}
\end{equation}

\subsection{Variable separation}

To simplify this equation, we make the substitution
\begin{equation}
g(z;\xi,E_{y}) = \exp \left( - i \omega z \right)
\tilde{g}(z;\xi,E_{y}),
\label{gtilde}
\end{equation}
where $\tilde{g}(s;\xi,E_{y})$ varies slowly comparing to 
$\exp \left( - i \omega z \right)$:
\begin{equation}
\partial \tilde{g} / \partial z \ll \omega \tilde{g}(z;\xi,E_{y}).
\label{slow}
\end{equation}
Equation (\ref{diffeqg}) takes the form 
\begin{equation}
 \frac{\partial \tilde{g}}{\partial z} 
\langle v_{z} \rangle
+
i \omega \tilde{g}(z;\xi,E_{y})
(1 - \langle v_{z} \rangle)
 = D_0 \left [ \frac{\partial }{\partial E_{y}}
\left(E_{y}
\frac{\partial \tilde{g} }{\partial E_{y}}
\right)
+
\frac{1}{E} 
\frac{\partial^2 \tilde{g} }{\partial \xi^2}
\right ] .
\label{diffeqgtilde}
\end{equation}
In the first term, the velocity can be approximated by unity:
$\langle v_{z} \rangle \approx 1$, i.e. the term
$
\partial \tilde{g} / \partial z 
( 1- \langle v_{z} \rangle )
$
can be neglected. However the term
$ i \omega \tilde{g}(z;\xi,E_{y})
(1 - \langle v_{z} \rangle) $
has to be kept because of (\ref{slow}).
Using the expression (\ref{v_z}) for $\langle v_{z} \rangle$, 
one obtains from (\ref{diffeqgtilde}) the 
following partial differential equation
for $\tilde{g}(z;\xi,E_{y})$
%:
%\begin{widetext}
\begin{eqnarray}
& & 
 \frac{\partial \tilde{g}(z;\xi,E_{y})}{\partial z} 
+
\frac{i \omega}{2 \gamma^2}  \tilde{g}(z;\xi,E_{y})
 = 
%& & \hspace*{2em}
D_0  \frac{\partial }{\partial E_{y}}
\left(E_{y}
\frac{\partial \tilde{g}(z;\xi,E_{y}) }{\partial E_{y}}
\right)
\label{diffeqgtilde1}
\\
& & 
\hspace*{5em}
- i \omega \frac{E_{y}}{2 E} \tilde{g}(z;\xi,E_{y})  
+ \frac{D_0}{E} 
\frac{\partial^2 \tilde{g}(z;\xi,E_{y}) }{\partial \xi^2}
- i \omega \frac{\xi^2}{2} \tilde{g}(z;\xi,E_{y})
\nonumber 
\end{eqnarray}
%\end{widetext}
This equation 
can be solved by the method of separation of 
variables. Putting
$\tilde{g}(z;\xi,E_{y}) = \mathcal{Z}(z) \Xi(\xi) \mathcal{E} (E_{y})$,
 after substitution into  (\ref{diffeqgtilde1})
we obtain a set of ordinary differential equations:
\begin{eqnarray}
\frac{D_0}{E} 
\frac{1}{\Xi(\xi)} 
\frac{d^2 \Xi(\xi)}{d \xi^2} 
- i \omega \frac{\xi^2}{2}
&=& \mathcal{C}_{\xi} , \label{eqXi} \\
\frac{D_0}{\mathcal{E}(E_{y})}
\frac{d }{d E_{y}}
\left(E_{y}
\frac{d  \mathcal{E}(E_{y})}{d E_{y}}
\right) 
- i \omega \frac{E_{y}}{2 E}
&=&  \mathcal{C}_{y}, \label{eqE} \\
\frac{1}{\mathcal{Z}(z)}
\frac{d \mathcal{Z}(z)}{d z} + 
\frac{i \omega}{2 \gamma^2} &=& \mathcal{C}_{z} , \label{eqZ} 
\end{eqnarray}
where $\mathcal{C}_{z}$, $\mathcal{C}_{\xi}$ and $\mathcal{C}_{y}$ do not
depend on any of the variables $z$, $\xi$ and $E_{y}$ and satisfy the condition 
\begin{equation}
\mathcal{C}_{z} = \mathcal{C}_{\xi} + \mathcal{C}_{y} .
\label{sumC}
\end{equation}

\subsection{Solving the equation for $\Xi(\xi)$}

Equation (\ref{eqXi}) can be rewritten as
\begin{equation}
\frac{d^2 \Xi(\xi)}{d \xi^2} 
- i  \frac{\omega E}{2 D_0} \xi^2 \Xi(\xi)
=  \frac{E}{D_0} \mathcal{C}_{\xi} \Xi(\xi). \label{eqXi1} 
\end{equation}
We change the variable
\begin{equation}
\chi = \mathrm{e}^{i \pi / 8} \sqrt[4]{\frac{\omega E}{2 D_0}} \; \xi
\end{equation}
and introduce the notation
\begin{equation}
\Omega = - \mathrm{e}^{- i \pi / 4} 
\sqrt{\frac{2 E}{\omega D_0}} \mathcal{C}_{\xi} .
\label{Omega}
\end{equation}
This results into 
\begin{equation}
\frac{d^2 \Xi(\chi)}{d \chi^2} 
-  \chi^2 \Xi(\chi)
= - \Omega \, \Xi(\chi). \label{eqXi2} 
\end{equation}
This equation has the form of the Schr{\"o}dinger equation for the harmonic oscillator.
Its eigenvalues and integrable eigenfunctions are well known:
\begin{eqnarray}
\Omega_{n} & = &  2 n + 1  \label{Omegan}\\
\Xi_{n}(\chi) & = & H_{n}(\chi) \exp(-\chi^2 / 2), \label{Xin}
\end{eqnarray}
where $n=0,1,2,\dots$ and 
$
H_{n}(\chi) = \mathrm{e}^{\chi^2} \left( - \frac{d}{d \chi} \right)^n \mathrm{e}^{-\chi^2}
$
are Hermite Polynomials
satisfying the orthogonality condition 
\begin{equation}
\int_{-\infty}^{+\infty} d \chi \; \mathrm{e}^{-\chi^2} H_{n}(\chi) H_{n'}(\chi) = 
\delta_{n n'} 2^{n} n! \sqrt{\pi}
\label{ortcondHn}
\end{equation}
which is equivalent to
\begin{equation}
\int_{-\infty}^{+\infty} d \chi \,  \Xi_{n}(\chi) \, \Xi_{n'}(\chi) = 
\delta_{n n'} 2^{n} n! \sqrt{\pi} .
\end{equation}
%. In particular, $H_{0}(\chi)  =   1$.

Returning back to the variable $\xi$ one obtains
\begin{equation}
\Xi_{n}(\xi) =  
H_{n} \left( \mathrm{e}^{i \pi / 8} \sqrt[4]{\frac{\omega E}{2 D_0}} \; \xi \right) 
\exp \left( -  \frac{1 + i}{4} \sqrt{\frac{\omega E}{D_0}} 
\xi^2 \right), 
\end{equation}

Any integrable function $F(\xi)$ can be represented as series
\begin{equation}
F(\xi) = 
F \left( \mathrm{e}^{-i \pi / 8} \sqrt[4]{\frac{2 D_0}{\omega E}} \chi \right)
=
\sum_{n=0}^{\infty} \mathfrak{b}_{n} \Xi_{n}(\chi)
\end{equation}
Let us multiply the above expression by $\Xi_{n'}(\chi)$ and integrate over $\chi$
\begin{equation}
\int_{-\infty}^{+\infty} d \chi
F \left( \mathrm{e}^{-i \pi / 8} \sqrt[4]{\frac{2 D_0}{\omega E}} \chi \right)
\Xi_{n'}(\chi) 
= \sum_{n=0}^{\infty} \mathfrak{b}_{n} 
\int_{-\infty}^{+\infty} d \chi
\Xi_{n}(\xi) \Xi_{n'}(\xi)
\end{equation}
Using (\ref{ortcondHn}) one finds  
\begin{equation}
\mathfrak{b}_{n} = 
\frac{1}{2^{n} n! \sqrt{\pi}}
\int_{-\infty}^{+\infty} d \chi \mathrm{e}^{-\chi^2 / 2}
F \left( \mathrm{e}^{-i \pi / 8} \sqrt[4]{\frac{2 D_0}{\omega E}} \chi \right)
H_{n}(\chi) 
\end{equation}

From (\ref{Omega}) and (\ref{Omegan}) one finds
\begin{equation}
\mathcal{C}_{\xi,n}
 = - (1+i)
\sqrt{\frac{\omega D_0}{E}} \left( n + \frac{1}{2} \right),
\ \ \ n = 0,1,2, \dots
\label{Cxi}
\end{equation}
%and
%\begin{equation}
%\Xi_{n}(\xi) =
%H_{n} \left( \mathrm{e}^{i \pi / 8} \sqrt[4]{\frac{\omega E}{2 D_0}} \; \xi \right) 
%\exp \left( -  \frac{1 + i}{4} \sqrt{\frac{\omega E}{D_0}} 
%\xi^2 \right).
%\label{Xin}
%\end{equation}
%Here
%$H_{n}(\dots)$ are Hermite polynomials.

\subsection{Solving the equation for $\mathcal{E}(E_{y})$}

Equation (\ref{eqE}) can be rewritten as 
\begin{equation}
\frac{d }{d E_{y}}
\left(E_{y}
\frac{d  \mathcal{E}(E_{y})}{d E_{y}}
\right) 
-  
\left(
\frac{i \omega}{2 D_0 E} E_{y}
+
\frac{\mathcal{C}_{y}}{D_0}
\right)
\mathcal{E}(E_{y})
=  0 , \label{eqE1}
\end{equation}
By the substitution 
\begin{equation}
E_{y} = \frac{1 - i}{2} \sqrt{\frac{D_0 E}{\omega}}
\varepsilon
\label{Eyvarepsilon}
\end{equation}
equation (\ref{eqE1}) can be reduced to
\begin{equation}
\varepsilon \frac{d^2 \mathcal{E}}{d \varepsilon^2} + \frac{d \mathcal{E}}{d \varepsilon} -
\left( \frac{\varepsilon}{4} - \frac{2 \nu + 1}{2} \right) \mathcal{E}  = 0
\label{eqLaguerreE}
\end{equation}
with 
\begin{equation}
2 \nu + 1
 = 
- (1 - i)
\sqrt{\frac{E}{\omega D_0}}
\mathcal{C}_{y} .
\label{TwoNuPlus1}
\end{equation}

Further substitution 
$\mathcal{E} (\varepsilon) = \exp(- \varepsilon / 2) L(\varepsilon) $
results into the Laguerre equation:
\begin{equation}
\varepsilon \frac{d^2 L}{d \varepsilon^2}
+ ( 1 - \varepsilon) \frac{d L}{d \varepsilon}
+ \nu L = 0
\end{equation}

One of two linearly independent solutions of this equation 
is logarithmically divergent at $\varepsilon \rightarrow 0$ and, therefore, has to be rejected. 
Another solution, $L_{\nu}(\varepsilon)$, is finite at $\varepsilon = 0$ and is known as 
the Laguerre function.\footnote{
At nonnegative integer
values of $\nu$, the Laguerre function is reduced to the well known Laguerre polynomials.
In the general case that is relevant to our consideration, it can be represented by 
an infinite series (\ref{Lnu_ser}).}

Returning back to the variable $E_{y}$, 
the solution of equation (\ref{eqE1}) can be represented as
%\begin{widetext}
\begin{equation}
\mathcal{E}(E_{y})
\! = \!
\exp \!
\left( \!
- \frac{1+i}{2} \sqrt{\frac{\omega}{D_{0} E}} E_{y}
\right)
L_{\nu} \! \!
\left( \!
(1+i) \sqrt{\frac{\omega}{D_{0} E}} E_{y} \!
\right)
\label{Ek}
\end{equation}

The eigenvalues can be found by imposing the boundary conditions.
If the energy of the channeling oscillations exceeds the value 
$U_{\max}$ (see Fig. \ref{potential.fig}) the particle enters
the region of high electron density, get scattered by crystal constituents
and becomes dechanneled.
Therefore, the distribution function of channeling particles
has to be zero at $E_y=U_{\max}$.
This results into the following boundary condition
\begin{equation}
L_{\nu} 
\left( 
(1+i) \sqrt{\frac{\omega}{D_{0} E}} U_{\max} 
\right) = 0.
\label{boundcond}
\end{equation}
Equation (\ref{boundcond}) has to be solved  for $\nu$.
Then, according to (\ref{TwoNuPlus1}), the eigenvalue 
$\mathcal{C}_{y,k}$ can be found from
\begin{equation}
\mathcal{C}_{y} = 
- \frac{(1+i)}{2} 
\sqrt{\frac{D_{0} \, \omega}{E}} (2 \nu + 1).
\label{Cynu}
\end{equation}
The subscript $k = 1,2,3,\dots$ enumerates different roots of
equation (\ref{boundcond}).

We introduce a dimensionless parameter 
\begin{equation}
\kappa = \frac{4}{j_{0,1}^2} \frac{\omega}{D_{0} \, E} U_\mathrm{max}^2
\label{kappa}
\end{equation}
($j_{0,k}$  is $k$-th zero of the 0-th order Bessel function: $J_0(j_{0,k})=0$).
Then equation (\ref{boundcond}) can be rewritten as 
\begin{equation}
L_{\nu} 
\left( \frac{1+i}{2} j_{0,1} \sqrt{\kappa} \right) = 0.
\label{eqnu}
\end{equation}
This equation has infinite number of complex roots  (see Appendix) which we denote as $\nu_{k}(\kappa)$, 
$k = 1,2,3,\dots$. The equation does not have any analytical solution and therefore has to be solved 
numerically.

Instead of the complex function $\nu_{k}(\kappa)$,  it is more convenient 
to introduce two real functions:
\begin{eqnarray}
\alpha_k (\kappa) & = & \frac{\sqrt{\kappa}}{j_{0,1}} 
\left[ 1 + 2 \left( \Re [ \nu_{k}(\kappa) ] - \Im [ \nu_{k}(\kappa) ] \right) \right]  
\label{alphak} \\
\beta_k (\kappa) & = & \frac{1}{2 j_{0,1} \sqrt{\kappa}} 
\left[ 1 + 2 \left( \Re  [ \nu_{k}(\kappa) ] + \Im [  \nu_{k}(\kappa) ] \right) \right]. 
\label{betak}
\end{eqnarray}
The eigenvalues (\ref{Cynu}) can be represented in the form
\begin{equation}
\mathcal{C}_{y,k} = - \frac{\alpha_k(\kappa)}{L_\mathrm{d}}  
- i \omega \theta_\mathrm{L}^2 \beta_k(\kappa).
\label{Cyk}
\end{equation}
Here 
\begin{eqnarray}
L_\mathrm{d} & = & 4 U_{\max} /(j_{0,1}^2 D_0) \ \  \mbox{and} \\
\theta_\mathrm{L} & = & \sqrt{2 U_{\max}/E}
\label{theta_L}
\end{eqnarray}
are, respectively, the dechanneling length \cite{BiryukovChesnokovKotovBook}
and Lindhard's angle. 
The parameter $\kappa$ (\ref{kappa}) can be rewritten in terms of $L_\mathrm{d}$ and $\theta_\mathrm{L}$
\begin{equation}
\kappa = \pi \frac{L_\mathrm{d}}{\lambda}  \theta_\mathrm{L}^2,
\label{kappa_Ld}
\end{equation}
where $\lambda = 2 \pi / \omega$ is the spatial period of the modulation.

\subsection{Solving the equation for $\mathcal{Z}(z)$}

Equation (\ref{eqZ}) has the solution 
\begin{equation}
\mathcal{Z}(z) = \exp \left(\mathcal{C}_{z} - i \frac{\omega}{2 \gamma^2} \right)
\label{Zz}
\end{equation}
The value of $\mathcal{C}_{z}$ can be found using (\ref{sumC}), (\ref{Cxi}) 
and (\ref{Cynu}). Then the solution (\ref{Zz}) takes the form
\begin{eqnarray}
\mathcal{Z}_{n,k}(z) & = & \exp \left \{
- \frac{z}{L_\mathrm{d}} 
\left[ \alpha_k(\kappa) + (2 n + 1) 
\frac{\sqrt{\kappa}}{j_{0,1}}
\right ] \right.
- \label{Znk} \\
& & 
\left.
i \omega z
\left [
\frac{1}{2 \gamma^2} + 
\theta_\mathrm{L}^2 \beta_k(\kappa) + 
\theta_\mathrm{L}^2  \frac{(2 n + 1)}{2 j_{0,1} \sqrt{\kappa}}
\right ]
\right \} . \nonumber
\end{eqnarray}
%\end{widetext}

Hence, the solution of Eq. (\ref{diffeqg}) is
represented as
\begin{equation}
g(z;\xi,E_{y}) = \exp \left( - i \omega z \right) \sum_{n=0}^{\infty} \sum_{k=1}^{\infty}
\mathfrak{a}_{n,k}
\Xi_{n}(\xi) \mathcal{E}_{k}(E_{y}) \mathcal{Z}_{n,k}(z) ,
\end{equation}
where the coefficients $\mathfrak{a}_{n,k}$ are found from the particle distribution at the
entrance of the crystal channel:
\begin{equation}
\mathfrak{a}_{n,k} = 
\frac{
\int_{-\infty}^{+\infty} d \xi
\int_{0}^{U_{\max}} d E_{y}
g(0;\xi,E_{y}) \Xi_{n}(\xi) \mathcal{E}_{k}(E_{y})
}{
2^n n! \sqrt{\pi} 
\int_{0}^{U_{\max}} d E_{y} 
\left [ \mathcal{E}_{k}(E_{y}) \right ]^2
}.
\end{equation}

\section{The demodulation length}

\subsection{The demodulation length in a straight channel}

Due to the exponential decrease of $\mathcal{Z}_{n,k}(z)$ with $z$
(see (\ref{Znk})), the asymptotic behaviour of $\tilde{g}(z;\xi,E_{y})$ 
at large $z$
is dominated by the term with $n=0$ and $k=1$
having the smallest value of the
factor $\left[ \alpha_k(\kappa) + (2 n + 1) \sqrt{\kappa}/j_{0,1}
\right ]$ in the exponential.
Therefore, at sufficiently large penetration depths, the particle
distribution depends on $z$ as
\begin{equation}
g (z;\xi,E_{y}) \propto \exp \left (
- z / L_{\mathrm{dm}}  - i \omega/u_{z} \, z
\right) 
\end{equation}
where $ L_{\mathrm{dm}}$ is the newly introduced parameter --- {\it the demodulation 
length}:
\begin{equation}
L_{\mathrm{dm}} = \frac{ L_\mathrm{d} }{\alpha_1(\kappa) + \sqrt{\kappa}/j_{0,1}}
\label{Ldm}
\end{equation}
and $u_{z}$ is the phase velocity of the modulated beam along the crystal channel
\begin{equation}
u_{z} = \left [ 1 + \frac{1}{2 \gamma^2} + 
\theta_\mathrm{L}^2 \left( \beta_k(\kappa) + \frac{1}{2 j_{0,1} \sqrt{\kappa}} \right) 
 \right  ]^{-1} .
\label{us}
\end{equation}
This parameter is important for establishing the resonance conditions between 
the undulator parameters and the radiation wavelength.

In this article we concentrate our attention on the demodulation length.
This parameter represents the characteristic scale of the penetration depth at
which a beam of channeling particles looses its modulation.

\begin{figure}[htb]
%\includegraphics*[width=12cm]{figure1.eps}
%%% for two column
\begin{center}
\includegraphics*[width=12cm]{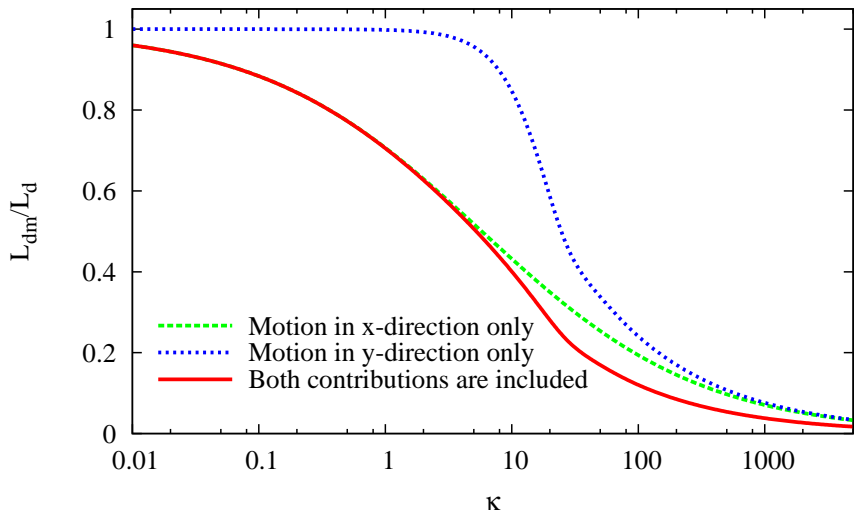}
\caption{The ratio of the demodulation length 
$L_{\mathrm{dm}}$ (\ref{Ldm})
to the dechanneling length $L_\mathrm{d}$ versus 
the parameter $\kappa$ (\ref{kappa_Ld}). See text for details.}
\end{center}
\label{Ldm.fig}
\end{figure}

Fig. \ref{Ldm.fig} presents the dependence of the ratio $L_{\mathrm{dm}}/L_\mathrm{d}$ on
the parameter $\kappa$. It is seen that the demodulation 
length approaches the
dechanneling length at $\kappa \lesssim 1$.
On the contrary, the ratio noticeably drops for $\kappa \gtrsim 10$.

It is instructive to study the influence of the particle motion in $x$ and $y$ direction
on the demodulation length separately. Replacing $\alpha_1(\kappa)$ in (\ref{Ldm}) with unity  
means neglecting the motion in the $y$ direction, while omitting the second term in the denominator
ignores the motion in $x$ direction. One sees from Fig. \ref{Ldm.fig} that
it is mostly the motion in $x$ direction that diminishes the demodulation length
at $\kappa \lesssim 10$,
while the influence of channeling oscillations is negligible.
This suggests the idea that for the axial channeling, i.e. when motion in both 
$x$ and $y$ directions has the nature of channeling oscillations, 
the demodulation length  $L_{\mathrm{dm}}$ may practically coincide with the dechanneling length $L_\mathrm{d}$ 
at higher frequencies of the beam modulation than in the case of planar channeling.

\subsection{The centrifugal force in a bent channel}

So far, beam demodulation in a straight channel has been considered. The channels of a crystalline 
undulator, however, have to be periodically bent. Therefore the above formalism has to be modified 
for the case of a bent channel.

Let us consider a crystal that is bent in the (yz) plane so that the crystal channel has a constant curvature
with the radius $R$. An ultrarelativistic particle with energy $E$ moving in such a channel 
experiences the action the centrifugal force
\begin{equation}
F_\mathrm{c.f.} = \frac{E}{R}.
\end{equation}

It is convenient to characterise the channel curvature by the dimensionless parameter $C$ defined 
as 
\begin{equation}
C = \left | \frac{F_\mathrm{c.f.}}{U'_\mathrm{max}}  \right |,
\end{equation}
where $U'_\mathrm{max}$ is the maximum value of the derivative of the particle potential energy 
in the channel, i.e. the maximum transverse force acting on the particle in the interplanar potential.
Channeling is possible at $0 \le C < 1$.
The value $C=0$ corresponds to a straight channel. The critical radius $R_c$ (known also as Tsyganov radius)
at which the interplanar potential becomes  unable to overcome the centrifugal force
corresponds to $C=1$.

In the case of potential energy (\ref{potential}), 
\begin{equation}
U'_\mathrm{max} = U'(\rho_\mathrm{max}) =
2  \frac{U_\mathrm{max}}{\rho_\mathrm{max}}.
\end{equation}
so that 
\begin{equation}
C= \frac{\rho_\mathrm{max} F_\mathrm{c.f.}}{2 U_\mathrm{max}}.
\end{equation}

\begin{figure}[ht]
%\includegraphics*[width=12cm]{figure1.eps}
%%% for two column
\begin{center}
\includegraphics*[width=10cm]{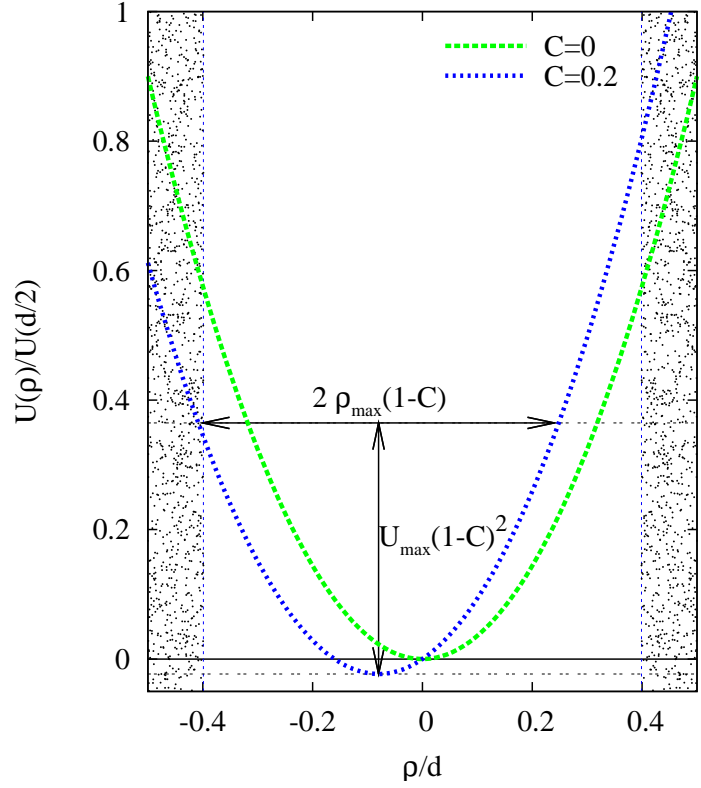}
\caption{The potential energy of a particle in the  planar crystal channel
for a straight, $C=0$, and for a bent, $C \not = 0$,  crystal.
The effective width of the bent channel is $2 \rho_\mathrm{max} (1-C)$ and the 
depth of the potential well is $U_\mathrm{max} (1-C)^2$,
where $2 \rho_\mathrm{max}$ and  $U_\mathrm{max}$ are, respectively, the effective
width and the depth of the straight channel (cf. Fig. \ref{potential.fig}).
}
\end{center}
\label{potential_C.fig}
\end{figure}

The potential energy is modified by the centrifugal force in the following way
\begin{equation}
U_{C}(\rho) = U(\rho) -  \rho F_\mathrm{c.f.}.
\end{equation}
For the parabolic potential energy (\ref{potential}) the modified potential can be 
conveniently rewritten in terms of the parameter $C$:
\begin{equation}
U_{C}(\rho) = U_\mathrm{max} 
\left [ 
\left( \frac{\rho}{\rho_\mathrm{max}} -C
\right)^2 -C^2
\right]
\end{equation}
The potential energy $U_{C}(\rho)$ reaches its minimum at $\rho_{0} = C \rho_\mathrm{max}$.
The effective width of the channel becomes (see Fig. \ref{potential_C.fig})
\begin{equation}
\rho_\mathrm{max} - \rho_{0} = \rho_\mathrm{max} (1-C).
\end{equation}
The depth of the potential energy well is 
\begin{equation}
U_{C}(\rho_\mathrm{max}) - U_{C}(\rho_{0}) = U_\mathrm{max} (1-C)^2.
\end{equation}
So to obtain the solution of the diffusion equation for the bent crystal we can use 
the results of Sec. \ref{solving} with the substitution
\begin{equation}
U_\mathrm{max}     \rightarrow  U_\mathrm{max} (1-C)^2 .
\label{subC}
\end{equation}

\subsection{The demodulation length in a bent channel}

Substitution (\ref{subC}) modifies the demodulation length and the Lindhard's angle
the parameter $\kappa$ in the following way:
\begin{eqnarray}
 L_\mathrm{d} & \rightarrow &  L_\mathrm{d}  (1-C)^2 \\
\theta_\mathrm{L} & \rightarrow &  \theta_\mathrm{L} (1-C)
\end{eqnarray}
Consequently, the the modification of parameter $\kappa$ is 
\begin{equation}
\kappa  \rightarrow \kappa (1-C)^4 .
\end{equation}
It is convenient to introduce modified functions $\alpha_k(\kappa, C)$ 
and $\beta_k(\kappa, C)$:
\begin{eqnarray}
\alpha_k(\kappa, C) & = & \frac{\alpha_k \left( \kappa (1-C)^4 \right)}{(1-C)^2} \\
\beta_k(\kappa, C)  & = & (1-C)^2 \beta_k \left( \kappa (1-C)^4 \right).
\end{eqnarray}
In terms of these functions, the eigenvalue $\mathcal{C}_{y,k}$ has the form 
\begin{equation}
\mathcal{C}_{y,k} = - \frac{\alpha_k(\kappa,C)}{L_\mathrm{d}}  
- i \omega \theta_\mathrm{L}^2 \beta_k(\kappa,C).
\label{CykC}
\end{equation}
This exactly coincides with (\ref{Cyk}) up to replacing $\alpha_k(\kappa)$
and $\beta_k(\kappa)$ with $\alpha_k(\kappa,C)$
and $\beta_k(\kappa,C)$, respectively. Note that $L_\mathrm{d}$ and $\theta_\mathrm{L}$
in (\ref{CykC}) have the same meaning as in (\ref{Cyk}): they  
are related to the straight channel.

Similarly, the demodulation length in the bent channel is given by 
\begin{equation}
L_{\mathrm{dm}} = \frac{ L_\mathrm{d} }{\alpha_1(\kappa,C) + \sqrt{\kappa}/j_{0,1}}
\label{LdmC}
\end{equation}

\begin{figure}[tbh]
%\includegraphics*[width=12cm]{figure1.eps}
%%% for two column
\begin{center}
\includegraphics*[width=12cm]{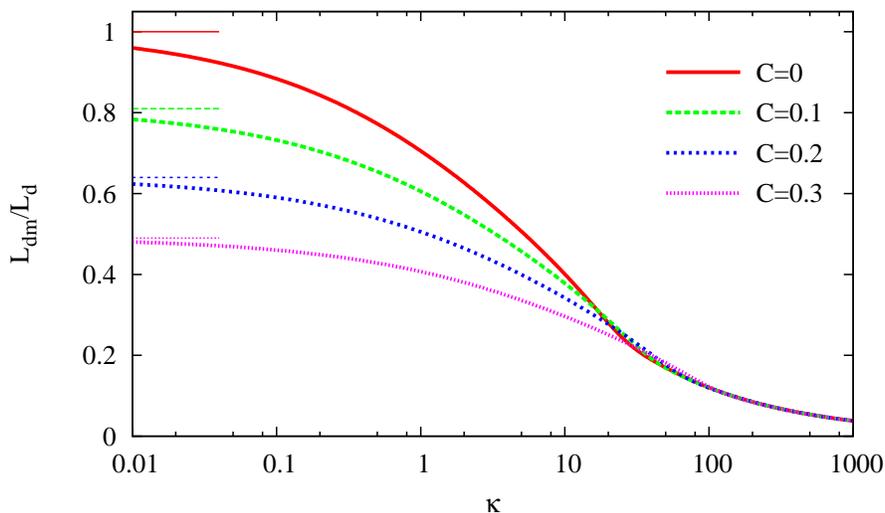}
\caption{The ratio of the demodulation length 
$L_{\mathrm{dm}}$ (\ref{LdmC})
to the dechannelling length in the straight channel $L_\mathrm{d}$ versus 
the parameter $\kappa$ (\ref{kappa_Ld}) for different values of curvature parameter $C$.
The corresponding asymptotic values at $\kappa \rightarrow 0$ are shown by thin 
horizontal lines.}
\end{center}
\label{Ldm_c.fig}
\end{figure}
Fig. \ref{Ldm.fig} presents the dependence of the ratio $L_{\mathrm{dm}}/L_\mathrm{d}$ on
the parameter $\kappa$. At $\kappa \rightarrow 0$, the demodulation length approaches 
$(1-C)^2 L_\mathrm{d}$ which is the dechannelling length in a bent crystal. It is seen that the demodulation 
length is smaller than dechannelling length by only 20--30\% at $\kappa \lesssim 1$ for $C$  ranging from
$0$ to  $0.3$. It noticeably drops, however, at $\kappa \gtrsim 10$.

The above estimations are made for the channel with constant curvature. Similar results are expected for 
a sinusoidal channel with $C$  varying between $0$ and  $0.3$.

It was proven for a number
crystals channels \cite{KSG2004_review}
that the dechannelling length of positrons  is sufficiently large to make the crystalline undulator feasible.
Such a crystalline undulator becomes a CUL, i.e. it generates coherent radiation, provided that it is
fed by a modulated positron beam and the beam preserves its modulation over the length of the crystal.
This takes place if the demodulation length in the crystalline undulator is not much smaller
than the dechannelling length. 
Hence, CUL is feasible if there exist crystal channels ensuring $\kappa \lesssim 1$
in the range of the photon energies above $\sim 100$ keV (softer photons are strongly absorbed
in the crystal). It will be shown in the next section that such crystal channels do exist.

\subsection{Estimation of the parameter $\kappa$}

To evaluate the parameter $\kappa$ (\ref{kappa_Ld}) we shall use the
approximate formula for the dechannelling length \cite{BiryukovChesnokovKotovBook}:
\begin{equation}
L_{d} =  \frac{256}{9 \pi^2} \frac{E}{m_\mathrm{e}} \frac{a_\mathrm{TF}}{r_{0}} \frac{d}{\Lambda}.
\label{Ld_BCK}
\end{equation} 
Here $m_\mathrm{e}$ and $r_{0}$ are, respectively, the electron mass and the classical radius, 
$d$ is the distance between the crystal planes,
and the Coulomb logarithm $\Lambda$ for positron projectiles is defined as \cite{KSG2004_review}:
\begin{equation}
\Lambda = \log \frac{\sqrt{2 E m_\mathrm{e}}}{I} - \frac{23}{24},
\label{Lambda}
\end{equation} 
with 
\begin{equation}
I \approx 16 Z^{0.9} \mbox{ eV}
\label{ion_pot}
\end{equation}
being the ionization potential of the crystal atom with the 
atomic number $Z$. The Thomas-Fermi radius of this atom is related to the Bohr radius 
$a_\mathrm{B}$ by the formula 
\begin{equation}
a_\mathrm{TF} = a_\mathrm{B} \frac{0.8853}{\sqrt[3]{Z}} .
\label{aTF}
\end{equation} 

Substituting (\ref{Ld_BCK}) into (\ref{kappa_Ld}) and taking into account 
(\ref{theta_L}), one obtains 
\begin{equation}
\kappa =  
\frac{512}{9 \pi} \frac{a_\mathrm{TF}}{\Lambda r_{0}} \frac{U_{\max}}{m_\mathrm{e}}  \frac{d}{\lambda}.
\label{kappa_BCK}
\end{equation} 
As is seen from the above formula the value of $\kappa$ is determined by the potential
depth $U_{\max}$, by the distance between the planes $d$ and the modulation period $\lambda$.
It also depends on the atomic number of the crystal atoms $Z$ via (\ref{ion_pot}) and (\ref{aTF}). 
These parameters are listed in table \ref{channels_table} for several crystal channels.
The dependence on the particle energy is cancelled out, except 
the weak dependence due to the logarithmic expression (\ref{Lambda}).

\begin{table}
\label{channels_table}
\begin{center}
\caption{The parameters of the crystalchannels used in the calculations (see text).
For (111) plane of diamond, only the larger of two channels is presented.}
\mbox{}\\
\begin{tabular}{|c|c|c|c|l|r|r|}
\hline \hline 
Crystal & $Z$ & $I$ (eV) & $a_\mathrm{TF}$  (\AA) & Plane & $d$ (\AA) & $U_{\max}$ (eV) \\
\hline \hline 
Diamond   &  6 &  80 &   0.26  & (100)   &   0.9 &   2.2 \\
          &    &     &         & (110)   &   1.3 &   7.3 \\
          &    &     &         & (111)L  &   1.5 &  10.8 \\
\hline
Graphite  &  6 &  80 &   0.26  & (0002) &   3.4 &  37.9 \\
\hline
Silicon   & 14 & 172 &   0.19  & (100)   &   1.4 &   6.6 \\
          &    &     &         & (110)   &   1.9 &  13.5 \\
\hline
Germanium & 32 & 362 &   0.15  & (100)   &   1.4 &  14.9 \\
\hline
Tungsten  & 74 & 770 &   0.11  & (100)   &   1.6 &  56.3 \\
\hline \hline
\end{tabular}
\end{center}
\end{table}

\begin{figure}[tbh]
%\includegraphics*[width=12cm]{figure1.eps}
%%% for two column
\begin{center}
\includegraphics*[width=12cm]{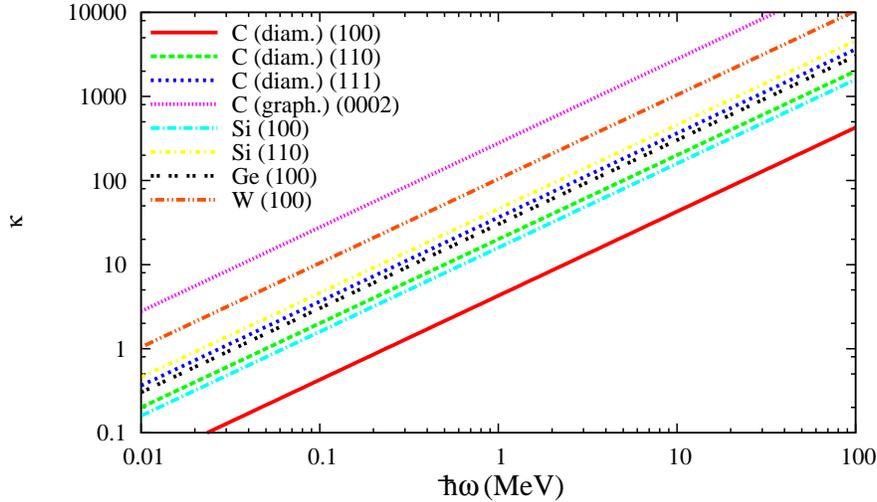}
\caption{The parameter $\kappa$ (\ref{kappa_Ld}) versus the photon energy
$\hbar \omega$ for the crystal channels listed in Table \ref{channels_table}.}
\end{center}
\label{hbaromega_kappa.fig}
\end{figure}

The dependence of the parameter $\kappa$ on the energy of the emitted photons, 
$\hbar \omega = 2 \pi \hbar/\lambda$, is shown in Fig. \ref{hbaromega_kappa.fig}.
The calculation was done for 1 GeV positrons.
Due to the weak (logarithmic) dependence of $\kappa$
on the particle energy,
changing this energy by an order of magnitude would leave
Fig. \ref{hbaromega_kappa.fig} practically unaltered.

As one sees from the figure,  $\kappa \sim 1$ corresponds to $\hbar \omega = 100-300$ keV
for (100) and (110) planes in Diamond and  (100) plane in Silicon. So these channels are 
the most suitable candidates for using in CUL. This is, however, not the case for a number of other crystals
e.g. for graphite and tungsten having $\kappa \gtrsim 10$ in the same photon energy range.

At $\hbar \omega \sim 10$ MeV, $\kappa$ becomes larger than $10$ for all crystal channels.
This puts the upper limit on the energies of the photons that can be generated by CUL.
It is expected to be most successful in the hundred keV range, while generating MeV photons
looks more challenging.

\section{Discussion and Conclusion}

One may expect that the demodulation is not limited to the processes illustrated in
figure \ref{demodulation.fig}. An additional contribution can come from the energy spread of the 
channelling particles, as it usually happens in ordinary FELs.
In fact, the contribution of the energy spread to the beam demodulation on the distance of
a few dechannelling lengths is negligible. It would be substantial if the relative 
{\it spread} $\delta E / E$ of particle energies
would be comparable to or larger than the ratio $\lambda_\mathrm{u}/L_{d}$. The latter 
ratio, however, can not be made smaller than $10^{-2}$ \cite{KSG2004_review},\footnote{Note that the 
corresponding quantity in ordinary ultraviolet and soft x-ray FELs, the inverse number of undulator periods 
$1/N_\mathrm{u} = \lambda_\mathrm{u}/L$, is usually of the order of $10^{-3}$--$10^{-4}$ \cite{SchmueserBook}.
That is why these FELs are so demanding to the small energy spread of the electron beam.}
while modern accelerators usually have a much smaller relative energy spread. The same is true for the energy spread
induced by the stochastic energy losses of the channelling particles due to the interaction 
with the crystal constituents and the radiation of photon. It was shown in Ref.\cite{KSG2000} that at initial
energies of $\sim 1 GeV$ or smaller, the average relative energy {\it losses} of a positron in the crystalline undulator 
$\Delta E / E$ are smaller than $10^{-2}$. Clearly, the induced energy {\it spread} $\delta E / E \ll \Delta E / E$
is safely below the ratio $\lambda_\mathrm{u}/L_{d}$.
From these reasons, we ignored energy spread of the particles in our calculations.

In conclusion, we have studied the propagation of a modulated positron beam in straight and bent planar
crystal channel within the diffusion approach and presented a detailed description of the used formalism.
We introduced a new parameter, the demodulation length, which characterizes the penetration depth 
at which the beam preserves its modulation.
It has been demonstrated that one can find the crystal channels where the demodulation length
sufficient for producing coherent radiation
with the photon energy of hundreds of keV. This opens the prospects for creating intense monochromatic 
radiation sources in a frequency range which is unattainable for conventional free electron lasers.
Developing suitable methods of beam modulation would be the next milestone on the way towards this goal.

\appendix
\numberwithin{equation}{section}
\section{Appendix: Solving equation (\ref{eqnu}).}

\subsection{A series expansion at $\kappa \ll 1$}

At small values of $\kappa$, the solution of equation (\ref{eqnu}) 
can be found in the form of power series.

The Laguerre function $L_{\nu}(\mathfrak{z})$ (which is a special case of the Kummer function, 
$
L_{\nu}(\mathfrak{z}) \equiv M (-\nu, 1, \mathfrak{z})
$
can be represented as 
\begin{equation}
L_{\nu}(\mathfrak{z})  = \exp(\mathfrak{z}/2) \sum_{n=0}^{\infty} A_{n}
 \left [ \frac{\mathfrak{z}}{2 (1+ 2 \nu)} \right ]^{n/2} 
 J_{n} \left ( \sqrt{2 (1 + 2 \nu) \mathfrak{z}} \right ),
\label{expansion}
\end{equation}
where $J_{n}(\dots)$ are Bessel functions and the coefficients $A_{n}$
are defined by the following recurrence relation:
\begin{eqnarray}
A_0 &=& 1 \\
A_1 &=& 0 \\
A_2 &=& \frac{1}{2} \\
A_{n+1} &=& \frac{1}{n+1} \left[ n A_{n-1} - (1 + 2 \nu) A_{n-2} \right ].
\end{eqnarray}

Keeping only the leading term in (\ref{expansion}) (this approximation is valid if $\mathfrak{z} \ll 1$), 
equation (\ref{eqnu})  can be reduced to
\begin{equation}
J_{0} \left ( \sqrt{(1+ 2 \nu^{(0)})  (1+i) j_{0,1} \sqrt{\kappa}   } \right ) = 0 ,
\label{eqJ0}
\end{equation}
where $\nu^{(0)}$ is the zero-order approximation to the root of equation (\ref{eqnu}).

Equation (\ref{eqJ0}) is satisfied if 
\begin{equation}
\sqrt{(1+ 2 \nu_{k}^{(0)})  (1+i) j_{0,1} \sqrt{\kappa}   } = j_{0,k},
\label{eq_j0k}
\end{equation}
$j_{0,k}$ is a root of the Bessel function: $J_{0}(j_{0,k})=0$.
Here the subscript $k=1,2,3,\dots$ enumerates the roots of the Bessel
function and the corresponding approximate solutions of equation (\ref{eqnu}).
Solving (\ref{eq_j0k}) for $\nu_{k}^{(0)}$ results into
\begin{equation}
\nu_{k}^{(0)}(\kappa) = \frac{1-i}{4} \frac{j_{0,k}^2}{j_{0,1}} \frac{1}{\sqrt{\kappa}}
- \frac{1}{2}.
\label{nu0}
\end{equation}

Keeping higher order terms in (\ref{expansion}) and expanding the Bessel functions around 
the the  zero-order approximation (\ref{nu0}),
one obtains a series expansion of $\nu_{k}(\kappa)$:
\begin{eqnarray}
\nu_{k}(\kappa) &=& \frac{1-i}{4} \frac{j_{0,k}^2}{j_{0,1}} \frac{1}{\sqrt{\kappa}}
-
\frac{1}{2} 
+
\frac{1+i}{24} j_{0,1}
\frac{j_{0,k}^2-2}{j_{0,k}^2} \sqrt{\kappa} 
\nonumber
\\
& & 
-
\frac{1-i}{720} j_{0,1}^3
\frac{j_{0,k}^4-17 j_{0,k}^2+54}{j_{0,k}^6}
\left ( \sqrt{\kappa} \right )^3 + \dots
\label{nuSeries}
\end{eqnarray}
(Dots stand for higher order terms with respect to $\kappa$).
For the functions (\ref{alphak}) and (\ref{betak}) expansion
(\ref{nuSeries}) takes the form
\begin{eqnarray}
\alpha_k(\kappa) & = & \frac{j_{0,k}^2}{j_{0,1}^2} 
-
 \frac{j_{0,1}^2 \left(j_{0,k}^4-17 j_{0,k}^2+54 \right)}{180 j_{0,k}^6}
\kappa^2
+ \dots 
\label{alphaSeries}
\\
\beta_k(\kappa)  &=& 
\frac{j_{0,k}^2-2}{12
   j_{0,k}^2}
+ \dots
\label{betaSeries}
\end{eqnarray}

\subsection{Numerical solution}
\begin{figure}[ht]
\begin{center}
\includegraphics*[width=10cm]{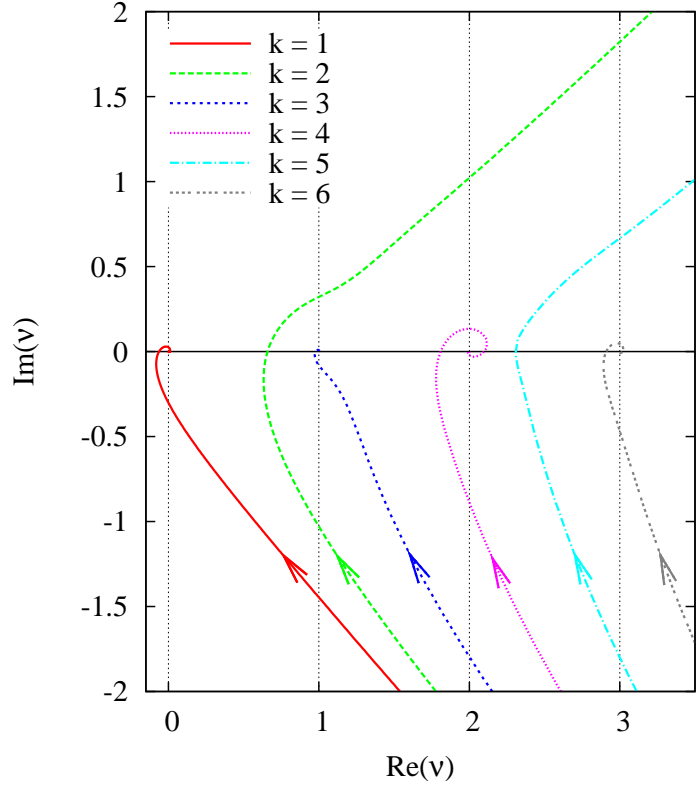}
\caption{The trajectories drawn by the roots $\nu_k(\kappa)$ of 
(\ref{eqnu}) in the complex plain at varying $\kappa$. The arrows show the direction of 
motion of the roots when $\kappa$ increases.}
\end{center}
\label{Plot_nu.fig}
\end{figure}

\begin{figure}[htb]
\begin{center}
\includegraphics*[width=10cm]{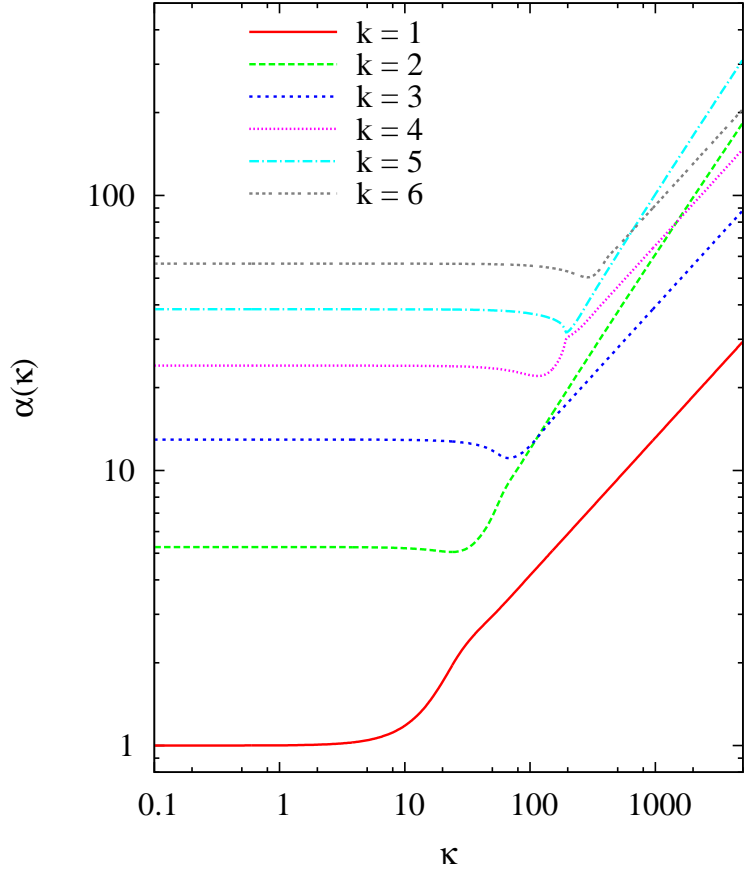}
\caption{The function $\alpha_k(\kappa)$ obtained by numerical analysis.}
\end{center}
\label{Plot_alpha.fig}
\end{figure}

\begin{figure}[hbt]
\begin{center}
\includegraphics*[width=10cm]{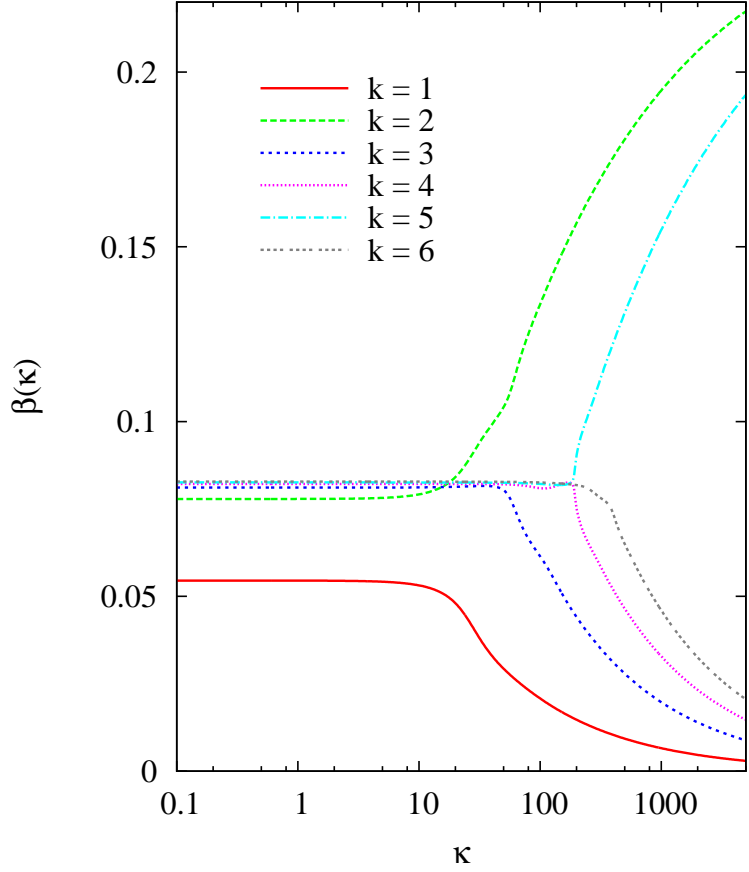}
\caption{The function $\beta_k(\kappa)$ obtained by numerical analysis.}
\end{center}
\label{Plot_beta.fig}
\end{figure}

Expansions (\ref{nuSeries}), (\ref{alphaSeries}) and (\ref{betaSeries}) fail at $\kappa \gtrsim 1$. 
Therefore, equation (\ref{eqnu}) has to be solved numerically.
In our numerical procedure,
we use the series representation for $L_{\nu}(\mathfrak{z})$:
\begin{equation}
L_{\nu}(\mathfrak{z}) =  \sum_{j=0}^{\infty}
 \frac{\prod_{m=0}^{j-1} (m - \nu)}{(j!)^2} \mathfrak{z}^j
\label{Lnu_ser}
\end{equation}
Equation was solved by Newton's method.
At small $\kappa$ the value found from the series expansion 
(\ref{nuSeries}) was used as initial approximation. Then 
$\kappa$ was gradually increasing. At each step, the equation
was solved and the solution was used as initial approximation for 
the next step. During this procedure, the roots $\nu_k(\kappa)$
were slowly moving in the complex plane along the trajectories 
shown  in figure \ref{Plot_nu.fig}.
The functions $\alpha_k$ and $\beta_k$ are plotted in Figures \ref{Plot_alpha.fig} and 
\ref{Plot_beta.fig}.

\subsection{Asymptotic behaviour at $\kappa \gg 1$}

As it is seen from Figure \ref{Plot_nu.fig}, some of $\nu_k(\kappa)$ approaches 
integer real numbers as $\kappa \rightarrow \infty$. This the case for 
$k=1,3,4,6$. For these solutions, the asymptotic behaviour can be found.

Let us represent $\nu_k(\kappa)$ in the form
\begin{equation}
\nu_k(\kappa) = n_{k} + \delta_k(\kappa),
\label{nuk_asymp}
\end{equation}
where $n_{k} = \nu_k(\infty)$ is an integer number and the function
$\delta_k(\kappa)$  goes to zero at $\kappa \rightarrow \infty$.

Substituting (\ref{nuk_asymp}) into (\ref{Lnu_ser}) and expanding
around $\delta_k=0$ one obtains
\begin{eqnarray}
L_{\nu}(\mathfrak{z}) &=& n_k!  \left [ 
\sum_{j=0}^{n_k} \frac{(-1)^{j}}{(j!)^{2} [n_k-j]!} \mathfrak{z}^{j}
\right.
\label{Lnu_asymp} \\
& & 
- \left. \delta_k 
\left (
P_{n_k} (\mathfrak{z})+
(-1)^{n_k} \mathfrak{z}^{n_k+1}
\sum_{j=0}^{\infty}
\frac{j!}{[(j+n_k+1)!]^{2}}
\mathfrak{z}^{j}
\right )
\right ]
\nonumber
\end{eqnarray}
Here $P_{n_k} (\mathfrak{z})$ is a polynomial of the order $n_k$
whose explicit form will not be needed in the following.

At $|\mathfrak{z}| \ll 1$, the infinite sum in (\ref{Lnu_asymp}) can be approximated by an
integral and evaluated by Laplace's method:
\begin{equation}
\sum_{j=0}^{\infty}
\frac{j!}{[(j+n_k+1)!]^{2}}
\mathfrak{z}^{j}
\asymp
\mathfrak{z}^{-2(n_k+1)}
\mathrm{e}^{\mathfrak{z}} 
\end{equation}
The polynomial $P_{n_k} (\mathfrak{z})$ in (\ref{Lnu_asymp}) becomes negligible with respect to 
the exponential at large $\mathfrak{z}$. Similarly, the leading order term dominates the first 
sum in (\ref{Lnu_asymp}). The asymptotic expression for $L_{\nu}(\mathfrak{z})$ takes, therefore, the following form 
\begin{equation}
L_{\nu}(\mathfrak{z}) \asymp (-1)^{n_k} n_k!  \left [ 
 \frac{\mathfrak{z}^{n_k}}{(n_k!)^{2} } 
-  
\delta_k \,
\mathfrak{z}^{-(n_k+1)}
\mathrm{e}^{\mathfrak{z}}
\right ]
\label{Lnu_asymp_z}
\end{equation}

Using (\ref{Lnu_asymp_z}) and taking into account (\ref{nuk_asymp}) one obtains the asymptotic expression
for the root of equation (\ref{eqnu}):
\begin{equation}
\nu_k(\kappa) \asymp n_{k} + \frac{1}{(n_k!)^{2}} \left( \frac{1+i}{2} j_{0,1} \sqrt{\kappa} \right)^{2 n_k+1}
\exp \left(- \frac{1+i}{2} j_{0,1} \sqrt{\kappa} \right).
\label{nuk_asymp_fin}
\end{equation}
This equivalent to the following asymptotic behaviour of the functions 
(\ref{alphak}) and (\ref{betak})
\begin{eqnarray}
\alpha_k(\kappa) & \asymp & 
\frac{2 n_{k} + 1}{j_{0,1}} \sqrt{\kappa} 
\label{alpha_asymp}
\\
& & 
+
 \frac{(j_{0,1})^{2 n_{k}} \kappa^{n_{k} + 1}}
{2^{n_{k} - 1} (n_k!)^{2}}
\exp \left ( - \frac{j_{0,1} \sqrt{\kappa}}{2} \right )
\sin \left( \frac{j_{0,1} \sqrt{\kappa}}{2} -\frac{\pi}{2} n_{k} \right)
\nonumber
\\
\beta_k(\kappa)  & \asymp & 
\frac{2 n_{k} + 1}{2 j_{0,1} \sqrt{\kappa}} 
\label{beta_asymp}
\\
& & 
+
 \frac{(j_{0,1})^{2 n_{k}} \kappa^{n_{k}}}
{2^{n_{k}} (n_k!)^{2}}
\exp \left ( - \frac{j_{0,1} \sqrt{\kappa}}{2} \right )
\cos \left( \frac{j_{0,1} \sqrt{\kappa}}{2} -\frac{\pi}{2} n_{k} \right)
\nonumber
\end{eqnarray}
It has to be stressed once more, that not all solutions of equation (\ref{eqnu}) have the above asymptotic 
behaviour. Among the solutions represented in figures \ref{Plot_nu.fig}-\ref{Plot_beta.fig}, 
(\ref{nuk_asymp_fin})--(\ref{beta_asymp}) is valid only for $k=1,3,4,6$ with $n_{k}=0,1,2,3$, respectively.

\section*{Acknowledgement}
This work has been supported in part by the European Commission 
(the PECU project, Contract No. 4916 (NEST)) and by
Deutsche Forschungsgemeinschaft.

\newpage
%%%%%%%%%% References

%%%%%%%%%%%%%%%%
\end{document}